# Watching lanthanide nanoparticles one at a time: characterization of their photoluminescence dynamics at the single nanoparticle level

Malavika Kayyil Veedu,[1] Gemma Lavilley,[2,3] Mohamadou Sy,[2] Joan Goetz,[2] Loïc J. Charbonnière,[3] Jérôme Wenger[1,*]

[1] Aix Marseille Univ, CNRS, Centrale Med, Institut Fresnel, AMUTech, 13013 Marseille, France

[2] Poly-Dtech, 204 avenue de Colmar, 67100 Strasbourg, France

[3] Equipe de Synthèse Pour l'Analyse, Institut Pluridisciplinaire Hubert Curien (IPHC), UMR 7178 CNRS/University of Strasbourg, Cedex 2, 67087 Strasbourg, France

*Corresponding author: jerome.wenger@fresnel.fr*

**Abstract:**

Lanthanide nanoparticles (LnNPs) feature sharp emission lines together with millisecond emission lifetimes which makes them promising luminescent probes for biosensing and bioimaging. Although LnNPs are gathering a large interest, their photoluminescence properties at the single nanoparticle level remain largely unexplored. Here, we employ fluorescence correlation spectroscopy (FCS) and photoluminescence burst analysis to investigate the photodynamics of Sm and Eu-based LnNPs with single nanoparticle sensitivity and microsecond resolution. By recording the photoluminescence intensity and the number of contributing LnNPs, we compute the photoluminescence brightness per individual nanoparticle, and estimate the actual number of emitting centers per nanoparticle. Our approach overcomes the challenges associated with ensemble-averaged techniques and provides insights into LnNP photodynamics. Moreover, we demonstrate our microscope capability to detect and analyze LnNPs at the single nanoparticle level, monitoring both photoluminescence brightness and burst duration. These findings expand our understanding of LnNPs and pave the way for their advanced biosensing applications at the single nanoparticle level.

**Introduction**

Lanthanide nanoparticles (LnNPs) are nanoscale particles composed of elements from the lanthanide series, a group of rare earth elements. Thanks to their unique set of photophysical characteristics showcasing sharp emission spectral peaks together with millisecond emission lifetimes, LnNPs have gathered a growing interest for advanced biosensing and bioimaging applications from immunoassays to multiplexed microscopies.[1,2] The slow emission dynamics of LnNPs act as both an advantage and a drawback. On one hand it allows to efficiently discriminate against the autofluorescence background by applying temporal gating on the detection, while on the other hand it implies a low radiative decay rate constant and hence a low photoluminescence (PL) emission rate per lanthanide emitter.[3,4] Fortunately, the development of LnNPs featuring a large number of emitting lanthanide centers embedded inside each nanoparticle and the so-called antenna effect taking advantage of chromophoric organic ligands to efficiently excite the lanthanide emitters have led to bright LnNPs sources overcoming the slow emission dynamics issues.[5,6]

While LnNPs are attracting a large interest, their photoluminescence properties remain primarily investigated through bulk ensemble-averaged spectroscopy. Detailed information at the single LnNP level is unfortunately seldom reported,[7] leaving several key questions unanswered. These challenges concern the possibility to detect LnNPs at the single nanoparticle level and to determine the effective brightness per LnNP, the number of emitting lanthanide centers per LnNP, and the role of the long photoluminescence lifetime in the detected signal. Answering these questions requires techniques from single-molecule fluorescence spectroscopy to achieve the proper resolution and sensitivity.[8,9] Within this toolbox, fluorescence correlation spectroscopy (FCS) provides a powerful approach to analyze the emission dynamics, unveiling the contributions of photon antibunching, triplet state blinking and translational diffusion.[10–12]

In this work, we use FCS and fluorescence burst analysis to investigate the photophysics of Eu and Sm doped TbNPs (respectively named as EuNP and SmNP) with single nanoparticle sensitivity and microsecond resolution. Nanoparticles doped with 1% of Eu or Sm coated with an organic ligand were prepared according to a previously published microwave assisted synthetic protocol.[6] PL burst analysis demonstrates that our experiment is able to detect and analyze a single LnNP, monitoring the PL brightness and PL burst duration at the single LnNP level as the nanoparticles diffuse across the confocal microscope detection volume. FCS data unveil the photoluminescence dynamics of SmNP and EuNP, demonstrating the absence of blinking at sub-millisecond timescales. FCS records the PL intensity together with the actual number of LnNPs contributing to the signal, allowing to compute the PL brightness per individual nanoparticle. The evolution of the brightness per LnNP as a function of the excitation power enables us to estimate the number of Eu and Sm emitting centers per nanoparticle, a quantity which is hard to obtain using other techniques. This work expands our understanding of LnNPs and paves the way for LnNPs applications using FCS and single molecule fluorescence techniques for advanced biosensing.

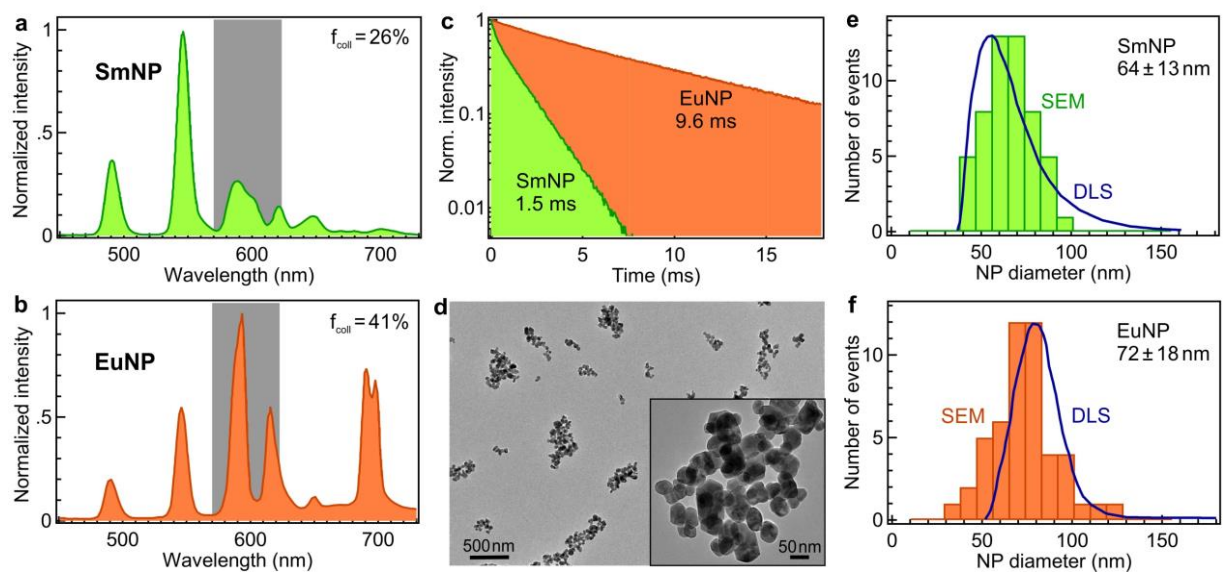

**Figure 1.** (a,b) Emission spectra of Sm and Eu doped TbNP ($\lambda_{exc}$ = 340 nm). The shaded area indicates the 570-620 nm region used for detection. (c) Photoluminescence lifetime decays of SmNP and EuNP. The intensity-averaged lifetime $\tau_{LT}$ is indicated on the graph. The individual components are 0.231 ms (6 %) & 1.52 ms (94 %) for SmNP and 2.74 ms (6 %) – 9.56 ms (94 %) for EuNP. (d) TEM image of EuNPs. (e,f) Diameter distributions for SmNP and EuNP obtained from SEM measurements (color bars) and DLS (blue line, scaled vertically). The average diameters and standard deviations as obtained by SEM are indicated on the graphs. These values are confirmed by TEM (Fig. S6).

## Results and Discussion

**Characterization of Sm and Eu nanoparticles**

The synthesized SmNP and EuNP feature distinctive and narrow emission bands which is a signature of lanthanide emission (Fig. 1a,b).[13] Both spectra feature emission bands typical of the Tb constituting the core with peaks at 485 and 545 nm originating from the $^5D_4 \rightarrow {}^7F_{6,5}$. For SmNP, the 570-620 nm region (shaded region, Fig. 1a), comprises a mixture of Tb emission bands ($^5D_4 \rightarrow {}^7F_{4,3}$), and a weak contribution arising from the $^4F_{9/2} \rightarrow {}^6H_{15/2}$ of Dy at ca 600 nm. In the case of EuNPs, this region is essentially corresponding to the two strong $^5D_0 \rightarrow {}^7F_{1,2}$ emission bands of Eu. Another specific characteristic of the LnNP emission is their millisecond excited state lifetime $\tau_{LT}$, corresponding to 1.52 ms for SmNP and 9.56 ms for EuNP (Fig. 1c). Experiments estimated the quantum yield $\phi$ of 72% for SmNP and 69% for EuNP. Luminescence quantum yields were measured according to conventional procedures,[14] with diluted solutions (optical density < 0.05), using rhodamine 6G in water ($\phi$ = 0.76)[15] and a Tb complex prepared in the laboratory (TbL(H$_2$O)Na, $\phi$ = 0.31)[16] as references. Errors on absolute quantum yields are estimated to be ± 15 %. Assuming a quantitative ligand to Ln energy transfer, these quantum yield numbers imply that upper limits for the radiative decay rate constants $k_{rad} = \phi/\tau_{LT}$ correspond to 0.47 ms$^{-1}$ for SmNP and 0.07 ms$^{-1}$ for EuNP. As compared to organic fluorescent dyes featuring $k_{rad}$ in the ns$^{-1}$ range,[17] the lanthanide emission is thus very slow, which would make the detection of a single rare earth atom very challenging in a aqueous solution at room temperature.[4] However the detection is eased by the fact that each LnNP contains a large number of emitting rare earth centers, one goal of this study being to determine experimentally this number.

The size distribution of the SmNPs and EuNPs is assessed by Transmission Electron Microscopy (TEM), Scanning Electron Microscopy (SEM) and Dynamic Light Scattering (DLS) (Fig. 1d-f). The diameter distributions determined by SEM and DLS are consistent with each other, indicating an average diameter of 64 ± 13 nm for SmNP and 72 ± 18 nm for EuNP (average ± standard deviation, Fig. 1e,f). With TEM, the average sizes are 56 ± 15 nm for SmNP and 64 ± 16 nm for EuNP, slightly smaller as compared to SEM (which we related to the better spatial resolution of TEM, see Fig. S6 and S7), yet with a difference much smaller than the standard deviation, so the TEM/SEM/DLS distributions can be considered as consistent to each other.

**Photoluminescence burst detection from single Sm and Eu nanoparticles**

Here we aim to detect single LnNPs as they diffuse across the confocal detection volume. The LnNP sample is diluted to a 50 pM concentration so as to reach an average number of particles in the confocal volume $N_{NP}$ around 0.05. As for single molecule fluorescence experiments,[18] this low number is a necessary condition to ensure that the bursts stem from a single nanoparticle. We set the excitation power to 5 µW for EuNPs so as to reach saturation while we benefit from the higher saturation power for SmNPs and use 20 µW to achieve a better signal to background ratio (more details about the excitation power dependence will be given in the next section). We take advantage of the long PL lifetime of LnNPs and apply temporal gating on the time-resolved detection to reject the background.[19,20] At 20 µW excitation power, nearly 85% of the total background intensity stem from Rayleigh and Raman scattering (see Supporting Information Fig. S1). As these processes are instantaneous and synchronized with the arrival of the laser excitation pulse, their contribution can be separated from the long millisecond lifetime of LnNPs. Here we apply a time gating on the detection channel [12,19] and select only the photons arriving within the interval 3 to 25 ns.

Figure 2a,b show subsections of the PL intensity time trace for SmNPs and EuNPs (the total trace duration is 600 s). PL peaks are clearly seen above the background, which we relate to the detection of a single LnNP crossing the detection volume. Each peak is analyzed to assess its brightness (in counts per ms) and its duration.[21] Figure 2c-e summarize our results on 400 bursts for each nanoparticle type. The SmNPs are brighter than the EuNPs, with a median brightness of 2.1 counts/ms as compared to 0.52 counts/ms (Fig. 3d). This higher brightness is related to the higher excitation power enabled by higher saturation power and shorter lifetime in the case of SmNPs as compared to EuNPs, as confirmed by FCS. The burst durations are shorter for SmNPs with a median value of 14.7 ms whereas EuNPs bursts have a median duration of 24 ms (Fig. 2e). These values stand in good agreement with the larger NP size (Fig. 1e,f) and longer FCS diffusion time of EuNPs as compared to SmNPs. We also find that the average burst duration is about 3 times the FCS diffusion time, which is expected in this field.[10]

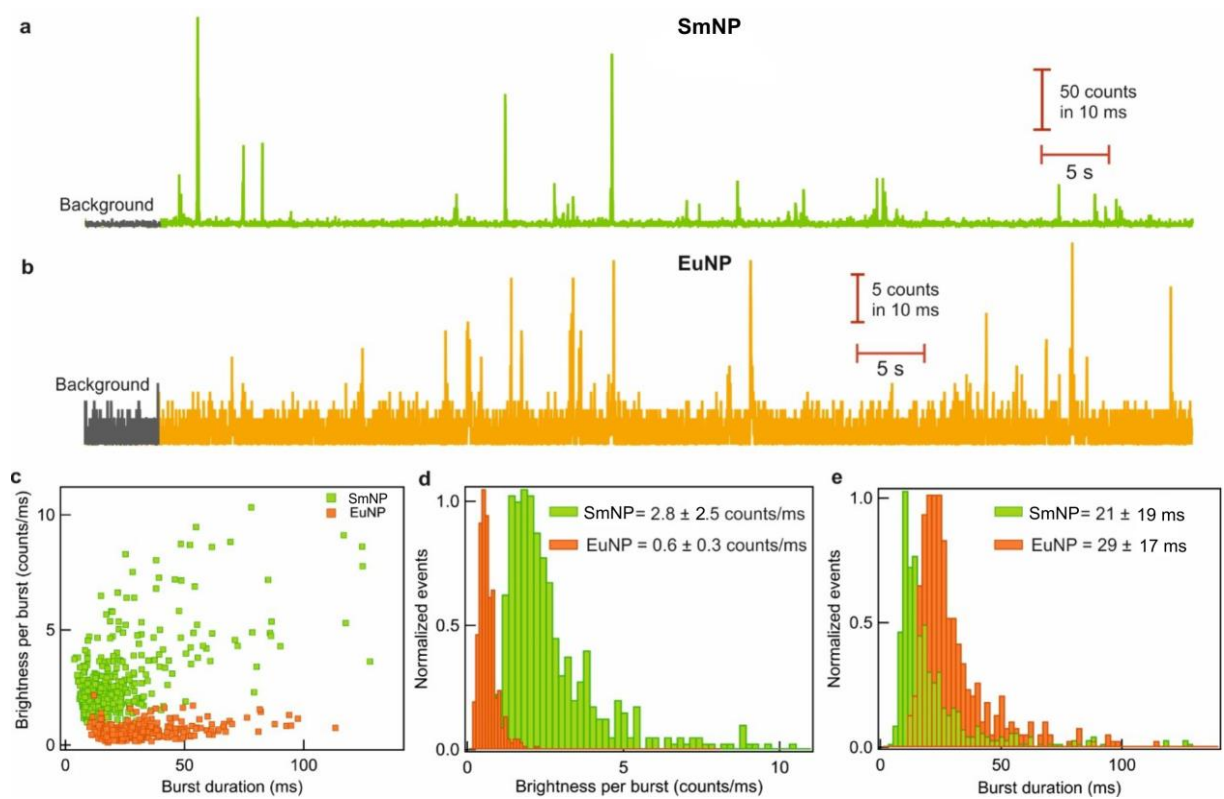

**Figure 2.** Demonstration of single lanthanide nanoparticle detection from photoluminescence burst analysis. (a, b) Example of photoluminescence time traces of SmNP and EuNP lanthanide nanoparticles at 50 pM concentration after time gating. The binning time is 10 ms. The laser power is 20 µW for (a) and 5 µW for (b). (c) Distribution of brightness per burst as a function of the burst duration. (d) Histogram of the brightness per burst of the lanthanide nanoparticles. (e) Histogram of the burst duration of the lanthanide nanoparticles. The acquisition time for these measurements were 10 min. The average values +/- one standard deviation are indicated on the graphs.

**FCS experiments on Sm and Eu nanoparticles**

FCS analyses the temporal fluctuations of the photoluminescence intensity stemming from the confocal detection volume. As for the PL burst analysis, we apply a time gating on the detection channel [12,19] and select only the photons arriving within the interval 3 to 25 ns (Fig. 3a). The background is reduced by 6.5 fold at 20 µW excitation power, down to less than 0.2 counts/ms (Fig. 3b & S1). At this low background level, the PL signal from the SmNPs and EuNPs clearly stands out (Fig. 3b), enabling a detailed FCS study.

Any change affecting the LnNP PL intensity (blinking, antibunching, translational diffusion) will affect the correlation data, modifying the FCS curve shape at specific timescales. The number of nanoparticles $N_{NP}$ within the confocal volume scales as the inverse of the FCS correlation amplitude and can be measured by FCS. Complete details about our FCS analysis and fitting model are given in the Materials & Methods section below.

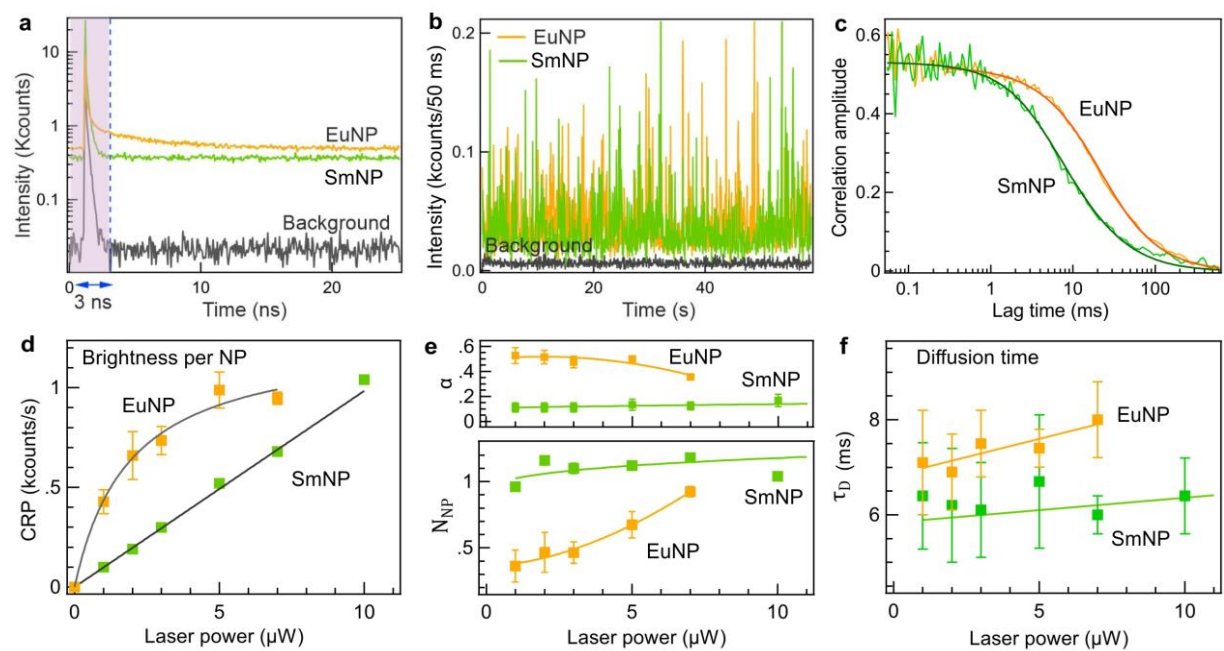

**Figure 3**. Fluorescence correlation spectroscopy characterization of lanthanide nanoparticles. (a) TCSCP decay of SmNP and EuNP showing time gating to discard the laser scattering contribution (purple area). (b) Intensity time trace of lanthanide nanoparticles after time gating. The average laser power was set to 5 µW. The concentration was 1 nM for SmNP and 0.4 nM for EuNP. (c) FCS correlation data (thin lines) for SmNP and EuNP at 7 µW excitation power. Thick color lines represent numerical fits using Eq. (5). (d-f) Comparison of different photophysical parameters of SmNP and EuNP lanthanide nanoparticles obtained from FCS data as a function of the excitation laser power. (d)

Photoluminescence brightness per nanoparticle, (e) Number of nanoparticles $N_{NP}$ in the confocal detection volume and amplitude factor α, (f) average diffusion time $\tau_D$. Error bars represent one standard deviation.

Figure 3c shows typical FCS correlation data for SmNPs and EuNPs at 7 µW excitation power. Supplementary correlation data for SmNP and EuNP at different excitation powers are given in Fig. S2. The FCS amplitude is clear, with a relatively low noise thanks to the diffusion time in the millisecond range enabling to accommodate for longer lag times and thus improve the signal-to-noise ratio in FCS.[10,22] The nearly flat shape at lag times below 1 ms indicates the absence of blinking on sub-millisecond timescales, as expected for a nanoparticle featuring several hundreds of rare-earth emitters. This is an advantage of LnNPs as compared to quantum dots or silver nanoclusters which are known for their significant blinking on a broad range of timescales.[23–26]

Figure 3d-f presents the results of the FCS analysis on both SmNPs and EuNPs with the different parameters extracted as a function of the excitation power: the brightness per nanoparticle $CRP$ (Fig. 3d), the number of nanoparticles in the confocal volume $N_{NP}$ (Fig. 3e), the translational diffusion time $\tau_D$ (Fig. 3f). The increase in the number of nanoparticles detected for the EuNPs with the laser power is an effect from saturation of the absorption-emission cycle, and has been reported earlier in FCS.[31] Looking at the FCS diffusion time (Fig. 3f), $\tau_D$ is slightly higher for EuNPs (7.4 ± 0.7 ms) than SmNPs (6.2 ± 0.6 ms) indicating that SmNPs are slightly smaller than EuNP, in agreement with the SEM and DLS characterization (Fig. 1e,f). Using a calibration of our microscope setup, we can compute the hydrodynamic radius $R_h$ of the nanoparticles using Stokes-Einstein equation, resulting in $R_h$ of 37 ± 4 nm for SmNPs and 44 ± 4 nm for EuNPs, again in good quantitative agreement with the DLS data (Fig. 1e,f). This highlights the capacity of FCS to simultaneously determine the size of the nanoparticles together with their photoemission dynamics (the influence of viscous media is presented in Fig. S4). As the LnNPs have a diffusion time in the millisecond timescale similar to the PL lifetime, we cannot exclude that some lanthanide emitters will emit their luminescence after the nanoparticle has exited the confocal detection volume. As a consequence, these photons will not be collected by our apparatus and the apparent brightness per nanoparticle will be lowered. Our characterization in Fig. 2d accounts for the amount of light collected on the detectors.

**From the brightness per nanoparticle to the number of emitters**

The brightness per nanoparticle $CRP$ (count rate per nanoparticle) is a key experimental parameter, as it mainly determines the outcome of any single nanoparticle experiment. The evolution of $CRP$ with the excitation power $P$ follows a typical saturation model (Fig. 3d):[8,17]

$$CRP = A \frac{P}{1+\frac{P}{P_{sat}}} \qquad (1)$$

Here the slope corresponds to

$$A = \eta\, f_{coll}\, \phi\, \sigma\, \rho\, N_{at}, \qquad (2)$$

where $\eta$ is the setup collection efficiency, $f_{coll}$ is the fraction of the PL spectrum emitted into the 570-620 nm detection window, $\phi$ is the quantum yield, $\sigma$ is the excitation cross-section, $\rho$ is the proportionality constant to accommodate the different units while expressing the excitation power $P$ in microwatts and $N_{at}$ is the number of lanthanide atoms per nanoparticle contributing to the detected signal. Using these notations, the saturation power of the nanoparticle is given by

$$P_{sat} = 1/(\tau_{LT}\, \sigma\, \rho) \qquad (3)$$

where $\tau_{LT}$ is the photoluminescence lifetime.

The key to assess the number of lanthanide emitters per nanoparticle $N_{at}$ is to compute the product $A\, P_{sat}$ which represents the brightness per nanoparticle in the saturation conditions. At high excitation powers exceeding $P_{sat}$, the brightness per nanoparticle reaches a saturation level of

$$A\, P_{sat} = \eta\, f_{coll}\, \phi\, N_{at}/\tau_{LT} = \eta\, f_{coll}\, k_{rad}\, N_{at} \qquad (4)$$

where $k_{rad} = \phi/\tau_{LT}$ is the radiative decay rate constant. This quantity no longer depends on the absorption cross section nor on the quantum yield. Only the overall collection efficiency, the radiative decay rate constant and the number of emitters per nanoparticle matter. Here, $A$ and $P_{sat}$ are determined experimentally from the FCS data (Fig. 1d). $f_{coll}$ can be easily computed from the PL spectra (Fig. 1a,b) while $k_{rad}$ is estimated by taking the ratio of the quantum yield and the PL lifetime. The setup collection efficiency $\eta$ is taken to be 5%, which is a typical value for confocal microscopes.[8,32]

For EuNPs, the analysis yields a slope $A$ of 660 counts.s$^{-1}$.µW$^{-1}$, while for SmNPs the slope is lower to 100 counts.s$^{-1}$.µW$^{-1}$. As both nanoparticles feature near similar quantum yields, the difference in their respective slopes $A$ is related to differences in the spectral collection efficiency $f_{coll}$ (41% for EuNPs and 26% for SmNPs) as well as in the excitation cross-section $\sigma$ and the number of lanthanide emitters per nanoparticle $N_{at}$. For EuNPs, we observe a clear saturation at 1.9 µW excitation power, which is consistent with the relatively long PL lifetime of 9.56 ms. For SmNPs, the saturation occurs beyond the range of laser powers available in our instrument, yet the analysis using the maximum power range available indicates a saturation power around 200 µW (Fig. S5). Finally, using Eq. (4), we estimate the number of emitter per nanoparticle to be 870 ± 120 for EuNPs and 3200 ± 1000 for SmNPs (the bigger uncertainty for SmNPs is related to the difficulty to measure $P_{sat}$). These number appear consistent with DRX experiments and the doping rate of emitters which indicate about 9000 Ln atoms for this type of NP.[5,6]

**Conclusions**

LnNPs are gathering a growing attention as novel luminescent probes for biosensing and bioimaging, yet their photoemission at the single nanoparticle level remained elusive as earlier works focused on ensemble-averaged techniques. Here, we use PL burst analysis and FCS to investigate the photoluminescence dynamic properties of LnNPs at the single nanoparticle level with sub-millisecond resolution. Contrarily to conventional colloidal semiconductor quantum dots[23–25] or DNA-encapsulated silver nanoclusters,[26] the SmNPs and EuNPs show negligeable blinking at sub-millisecond timescales. Recording the brightness per individual nanoparticle links the occurrence of photosaturation with the PL lifetime. It also enables the estimation of the number of emitting centers per nanoparticle, a quantity which is challenging to obtain using other techniques. Our microscope system is able to resolve individual PL bursts stemming from a single Sm or Eu doped nanoparticle, allowing to record the PL brightness and burst duration at the single LnNP level. Gaining new insights on the LnNP photodynamics and demonstrating single LnNP detection are significant steps forward to diversify the use of LnNPs in biosensing applications.

**Materials and Methods**

**Sample preparation.** Nanoparticles doped with 1% of Eu or Sm coated with an organic ligand were prepared according to a previously published microwave assisted synthetic protocol.[6] The nanoparticles were sonicated for 20 min before experiments to prevent aggregation, then diluted in Milli-Q water for the FCS experiments.

**Electron microscopy.** The SEM images were recorded on a FEI DB235 Strata electron microscope equipped with field emission gun (FEG) source. The TEM images were recorded on a JEOL 2100F electron microscope operating at 200kV equipped with a GATAN GIF 200 electron imaging filter. Powder samples were dispersed in water and a drop of this suspension deposited on TEM grids. The grid was prepared with a porous membrane covered by an amorphous carbon layer. In order to avoid disturbing random signal coming from amorphous carbon, detected LnF$_3$ particles were those which lies on strand of these holes. The images produced were analysed using *ImageJ* software. Core diameters were measured on samples comprising around 100 nanoparticles. The straight-line tool was used to measure the distance from one edge of the nanoparticles to the other.

**Optical microscopy.** All the FCS measurements are done in a home-built confocal microscope setup. The nanoparticles are excited at 490 nm by an iChrome-TVIS laser (Toptica GmbH, pulse duration ~3 ps) with a repetition rate of 40 MHz. To reflect the laser towards the microscope we use a multiband dichroic mirror (ZT 405/488/561/640rpc, Chroma). The excitation light is focused by a Zeiss C-Apochromat 63×, 1.2 NA water immersion objective lens. The fluorescence signal is collected by the same objective lens in an epifluorescence configuration and then passes through the same multiband dichroic mirror. An emission filter (ZET405/488/565/640mv2, Chroma) is used to block the laser back reflection and an 80 µm confocal pinhole is used to select the fluorescence signal. The emitted photons in the 570-620 nm spectral range are recorded by an avalanche photodiode APD (Picoquant MPD-5CTC). The photodiode output is connected to a time-correlated single photon counting (TCSPC) module (HydraHarp 400, Picoquant) and each fluorescence photon is recorded with individual timing and channel information in a time-tagged time-resolved (TTTR) mode. The DLS data was recorded on on an AMERIGO™ particle size & zeta pontential analyzer from Cordouan Technologies, equipped with a VASCO KIN™ particle size analyzer for the measurements of DLS diameters.

**Single Nanoparticle Burst Analysis.** We used the Burst Analysis module of the PIE Analysis with MATLAB (PAM) for this work.[21] All Photon Burst Search function was used to perform burst detection where we input the threshold photon counts, the time window of the burst, and the total photon counts per burst. Analysis of SmNP (EuNP) was done considering each peak as a single molecule burst having at least 20 (20) photons per burst and a minimum of 2 (2) photons per burst time window of 2 (4) ms.

**FCS Analysis.** FCS traces are obtained from the auto-correlation of the fluorescence intensity time trace from the APD. We time-gated and discarded photons within the 3 ns window centered on the laser pulse arrival time to reduce the background due to laser-induced scattering and Raman scattering. The FCS integration time was 300 s for laser powers below 3 µW, 240 s for other cases, except for SmNP at excitation power higher than 7 µW where 120 s integration time was found sufficient. To fit the FCS curves, we used the following model:[10]

$$G(\tau) = \left(1 - \frac{B}{F}\right)^2 \frac{1}{N_{NP}} \left(1 - \alpha\, e^{\left(\frac{-\tau}{\tau_A}\right)}\right)\left(1 + \frac{\tau}{\tau_D}\right)^{-1} \left(1 + \frac{\tau}{\kappa^2 \tau_D}\right)^{-0.5} \quad (5)$$

Where $G(\tau)$ is the auto-correlation function at time $\tau$, $N_{NP}$ is the total number of nanoparticles in the confocal volume, $\alpha$ is the amplitude of the correction term needed to correctly interpolate the FCS data in the millisecond timescale. We relate its origin to the millisecond lifetime of the LnNPs (see Supporting Information Fig. S3 and S4). While we presently do not have a solid theoretical foundation for this correction term, it should be noted that it remains a minor correction and it does not significantly modifies our findings and main scientific conclusions. $\tau_D$ is the diffusion time and $\tau_A$ is set to correspond to the lifetime of NPs (1.52 ms for SmNPs and 9.56 ms for EuNPs). $\kappa$ corresponds to the aspect ratio of the axial to the transversal dimension of the detection volume, set to $\kappa$ = 5 based on our past results.[33] $B$ is the measured background for confocal experiments which typically amounts to 195 counts at 20 µW excitation power (Fig. S1). $F$ is the total detected fluorescence intensity. We calculate the brightness per nanoparticle $CRP$ using

$$CRP = \frac{1}{N_{NP}}(F - B) \quad (6)$$

## Supporting Information

Background intensity after time gating, Supplementary correlation data for SmNPs and EuNPs at different excitation powers, Fit residuals for different models, SmNP correlation data in viscous media, Evolution of SmNP brightness as a function of laser power, TEM and SEM images of SmNPs and EuNPs, ICP-AES analysis of the samples.

## Acknowledgments

This project has received funding from the European Research Executive Agency (REA) under the Marie Skłodowska-Curie Actions doctoral network program (grant agreement No 101072818).

## Conflict of Interest

The authors declare no conflict of interest.

## Data Availability Statement

The data that support the findings of this study data are available from the corresponding author upon request.

**Supporting Information for**

**Watching lanthanide nanoparticles one at a time: characterization of their photoluminescence dynamics at the single nanoparticle level**

Malavika Kayyil Veedu,[1] Gemma Lavilley,[2,3] Mohamadou Sy,[2] Joan Goetz,[2] Loïc J. Charbonnière,[3] Jérôme Wenger[1,]*

[1] Aix Marseille Univ, CNRS, Centrale Med, Institut Fresnel, AMUTech, 13013 Marseille, France

[2] Poly-Dtech, 204 avenue de Colmar, 67100 Strasbourg, France

[3] Equipe de Synthèse Pour l'Analyse, Institut Pluridisciplinaire Hubert Curien (IPHC), UMR 7178 CNRS/University of Strasbourg, Cedex 2, 67087 Strasbourg, France

*Corresponding author: jerome.wenger@fresnel.fr*

**Contents:**

## S1. Background intensity after time gating

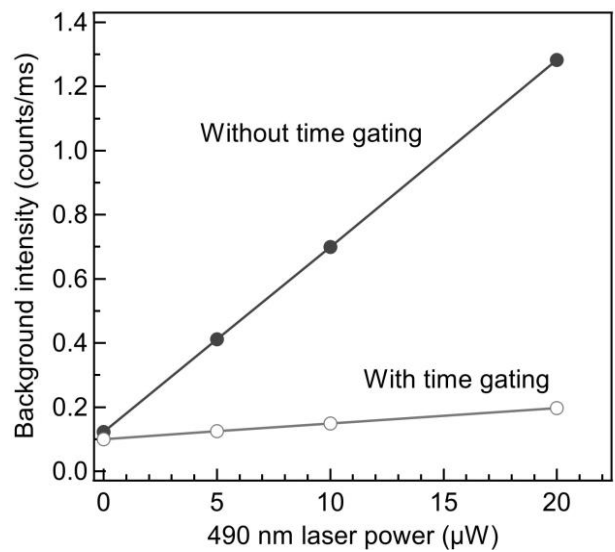

**Figure S1.** Evolution of the total background intensity recorded on a water solution without the LnNPs as a function of the 490 nm laser power. Filled markers indicate the experiment where all photons are counted, without applying any temporal gating. Empty markers show the background intensity reduction after applying a temporal gating filter where all photons incoming within the 0-3 ns interval are discarded for the analysis, as illustrated on the TCSPC data in Fig. 3a. Lines are linear fits to the data.

**S2. Supplementary correlation data for SmNPs and EuNPs at different excitation powers**

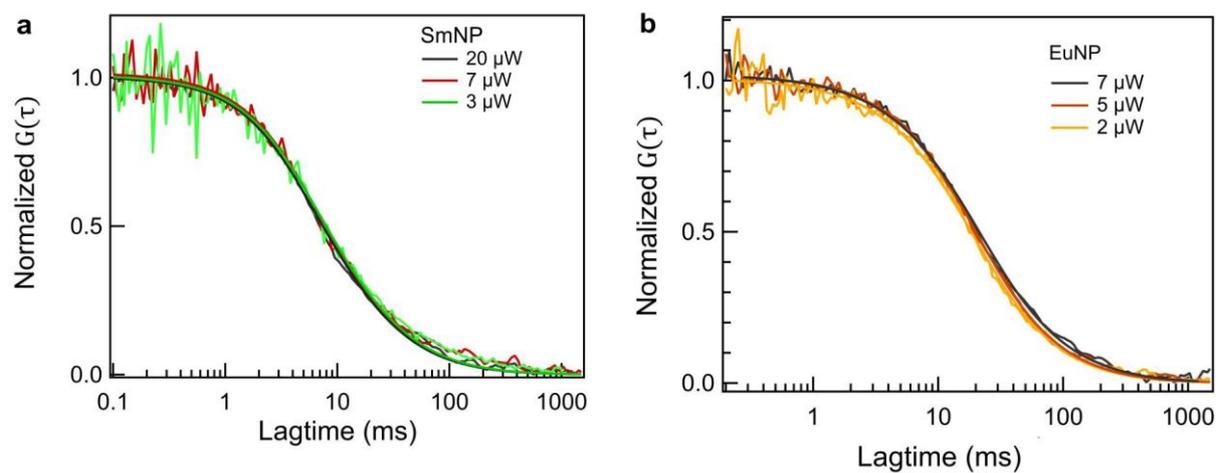

**Figure S2.** FCS correlation data for SmNPs (a) and EuNPs (b) at increasing excitation powers show a remarkable consistency in the correlation shape indicating no sign for blinking nor photobleaching.

## S3. Fit residuals for different models

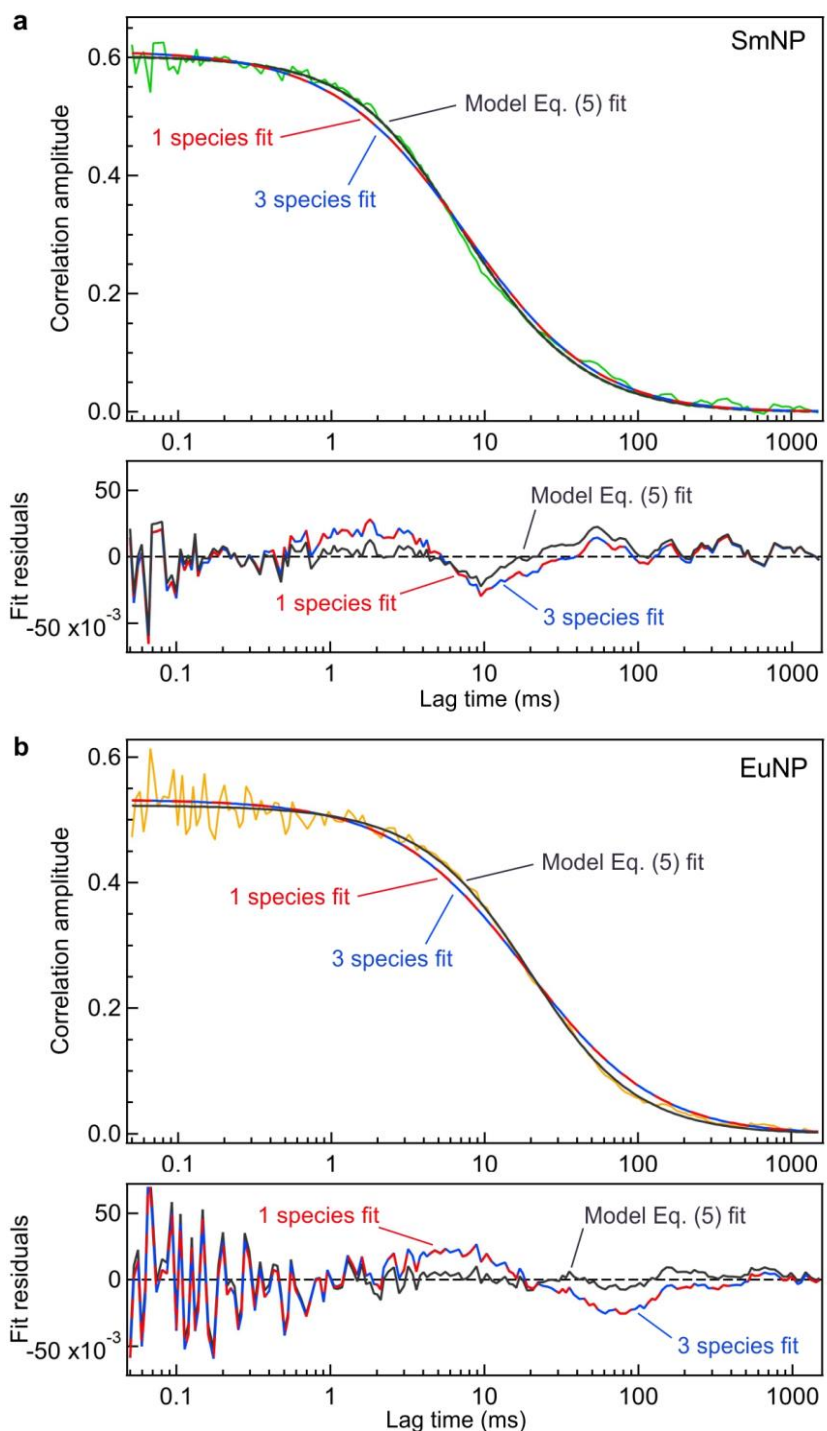

**Figure S3.** FCS correlation data for SmNPs (a) and EuNPs (b) together with the numerical fits and the fit residuals (lower graph). The fit using a single species (red trace) nearly perfectly overlaps with the fit using 3 species (blue trace). For the 3 species model, we set the relative diffusion times to follow the size histogram derived in Fig. 1, with diffusion times 0.75×, 1× and 1.25× the average diffusion time obtained from the single species fit. This led to no improvement in the fit residuals. Fitting with a 3 species model where all fit parameters are free led to an identical result as the 1 and 3 species fits shown here. The laser power was 20 µW for SmNP and 5 µW for EuNP.

## S4. SmNP correlation data in viscous media

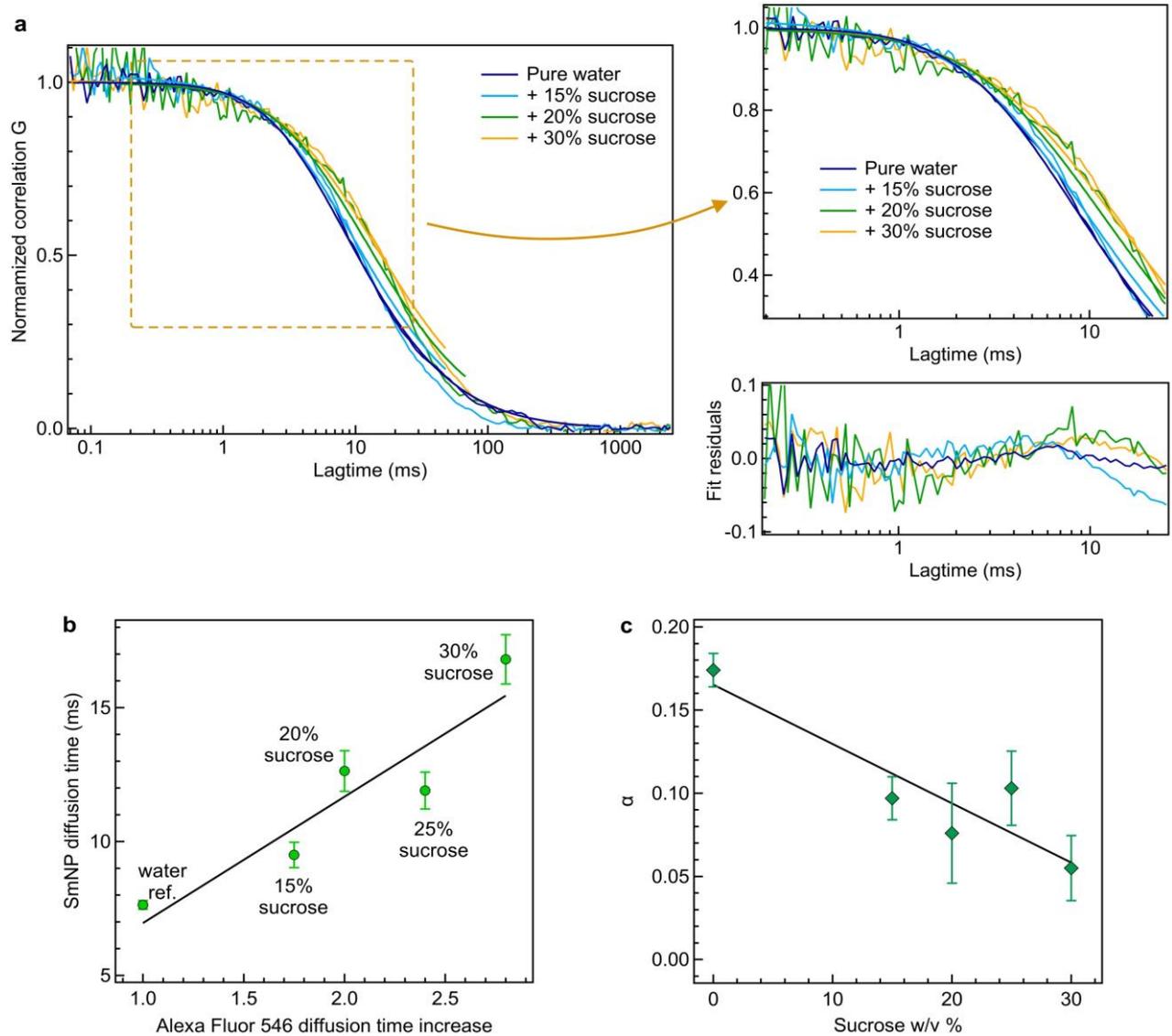

**Figure S4.** (a) FCS correlation data (thin lines) for SmNP in solutions containing different amounts of sucrose (in weight/volume) to increase the dynamic viscosity. The experimental conditions are similar to Fig. 3c, S2 and S3, with the excitation power being set to 20 µW. The thick color lines represent numerical fits using Eq. (5), where $\tau_A$ is set to correspond to the SmNP lifetime. (b) SmNP diffusion time $\tau_D$ for different sucrose concentrations. The influence of sucrose on the increase in the apparent FCS diffusion time was calibrated using standard FCS performed on Alexa Fluor 546 at 10 nM final concentration with the same microscope. We note a linear correlation (Pearson's correlation coefficient 92.5%) between the increase in the SmNP diffusion time $\tau_D$ and the increase in the diffusion time noted for Alexa Fluor 546. (c) Correction amplitude $\alpha$ using the model in Eq. (5) for different sucrose concentrations. The amplitude decreases with higher amounts of sucrose (Pearson's correlation coefficient -91.3%) but still remains clearly positive. These findings further substantiate the validity of our approach.

## S5. Evolution of SmNP brightness as a function of laser power

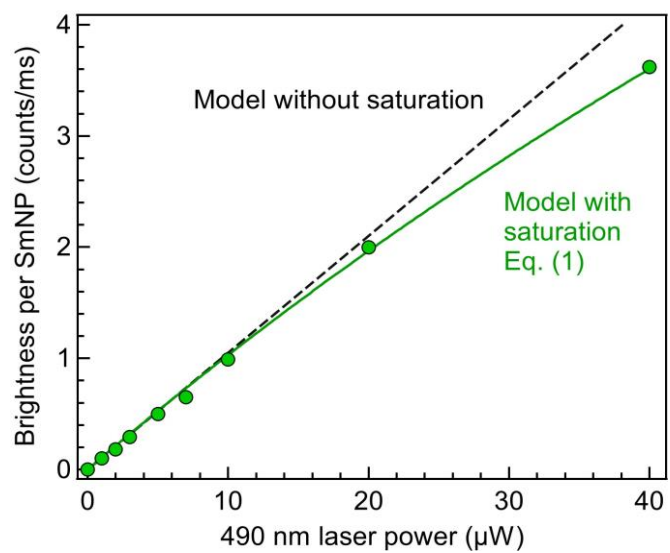

**Figure S5.** Evolution of the brightness per nanoparticle CRP as a function of the 490 nm laser power. The data is similar to Fig. 3d but on a broader power range, limited by the maximum laser power of our instrument at this illumination wavelength. Filled markers are experimental data points. The dashed line is a linear fit to the data, while the green line takes into account the saturation evolution based on Eq. (1). Using this data, we can estimate a saturation power around 200 µW for SmNPs.

**S6. Transmission electron microscope and SEM images of SmNPs and EuNPs**

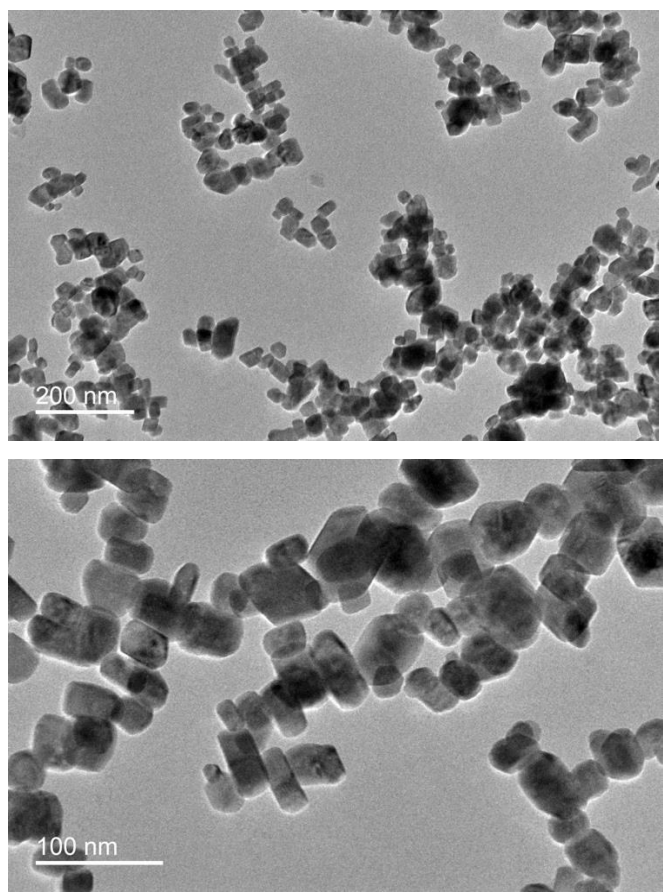

**Figure S6.** TEM images of SmNPs (La0.14Tb0.85Sm0.1F3). TEM revealed the presence of nanoparticles with an average diameter of 56 ± 15 nm, with a clearly elongated structure for the smallest particles. Transmission Electron Microscopy (TEM) was performed with a JEOL 2100F electron microscope operating at 200kV equipped with a GATAN GIF 200 electron imaging filter.

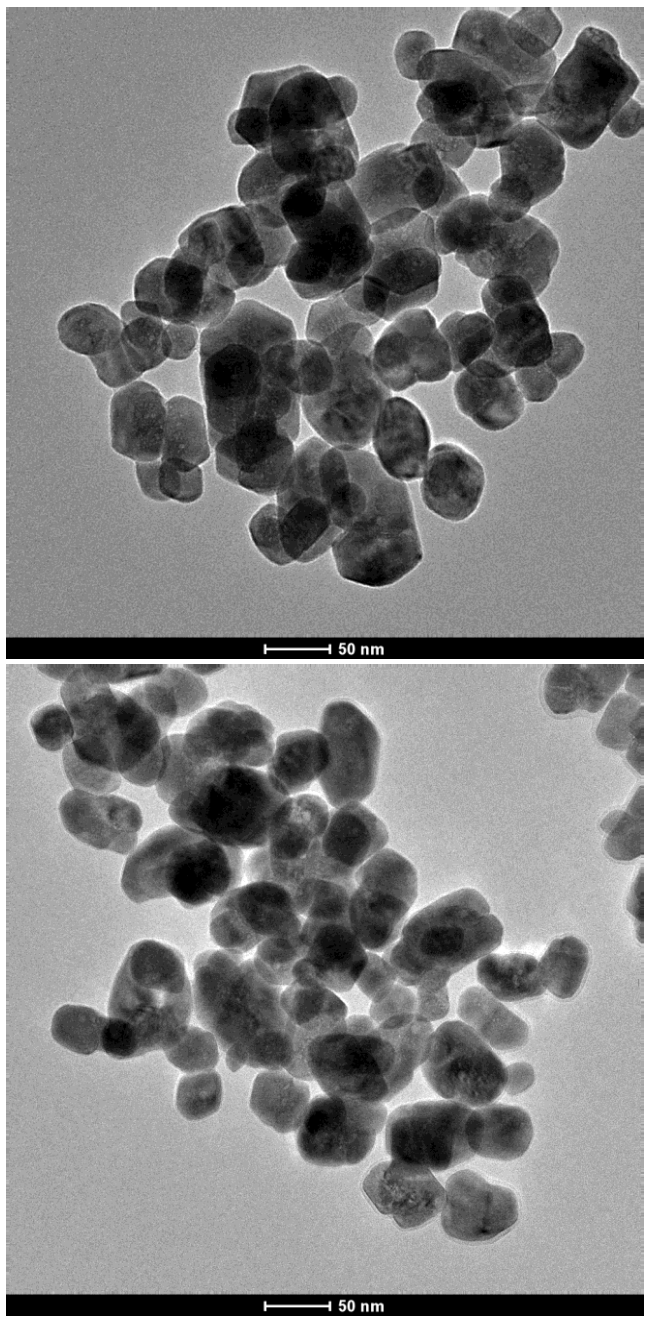

**Figure S7.** TEM images of EuNPs

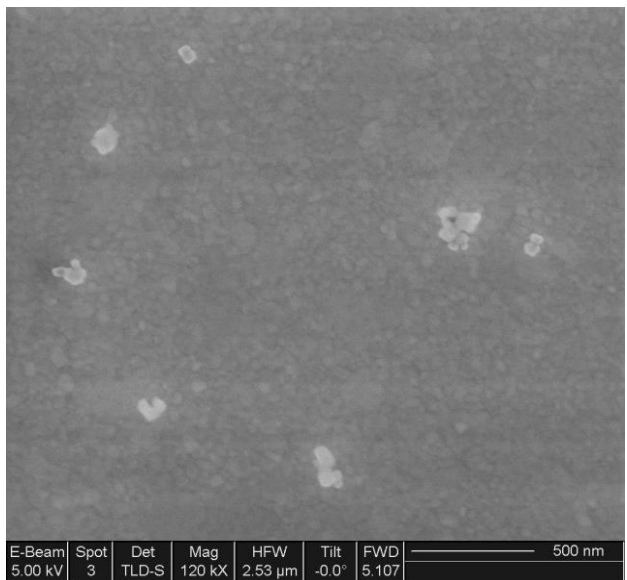
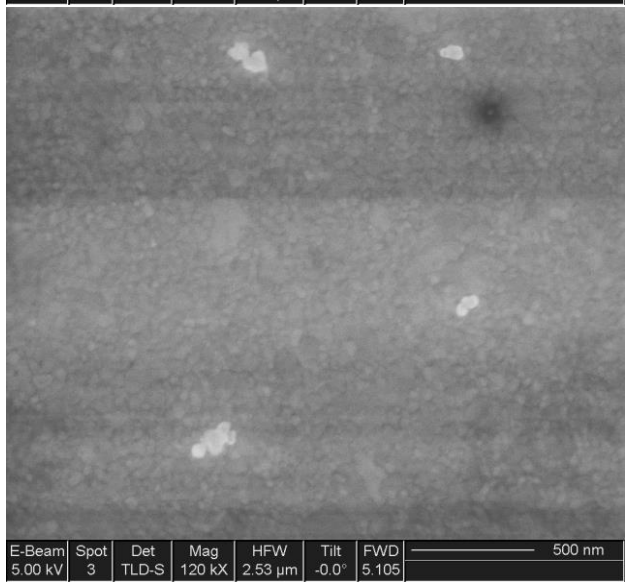

**Figure S8.** SEM images of SmNPs

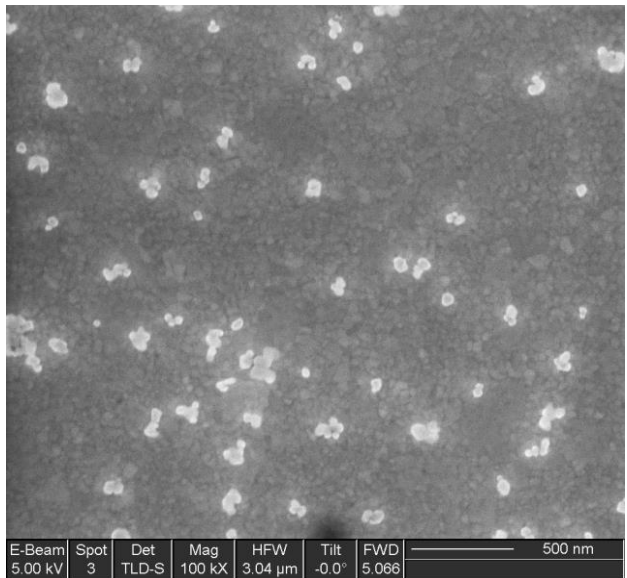
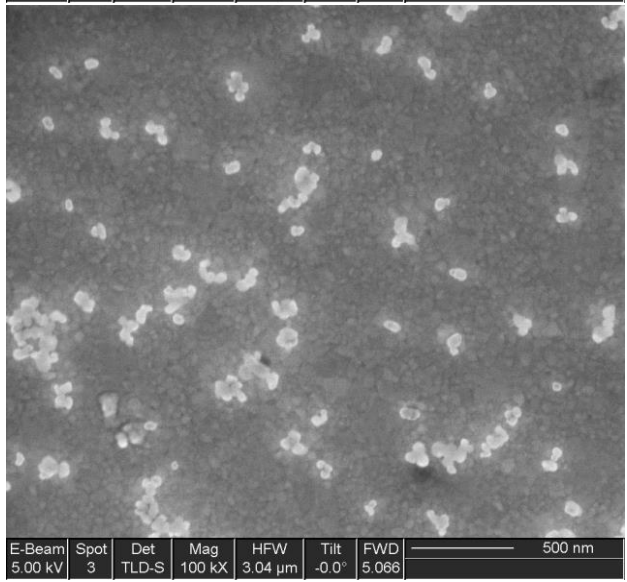

**Figure S9.** SEM images of EuNPs

## S7. Detailed analysis of the nanoparticle

| Type | DLS (nm) | TEM (nm) | ICPAES results (mg/kg) | | |
|---|---|---|---|---|---|
| | | | La | Sm | Tb |
| Nps Tb/Sm/La 85%/1%/14% | 78 | 56 | 490 | 35 | 3116 |
| Nps Tb/Eu/La 85%/1%/14% | 83 | 64 | 540 | 49.3 | 4327 |

**Table S1.** Inductively Coupled Plasma Atomic Emission Spectroscopy (ICP-AES) analysis of samples in water were performed with a Varian 720 spectrometer equipped with a quartz Meinhard nebulizer and a cyclone spray chamber. Anne Boos and Pascale Ronot are gratefully acknowledged for performing the ICP/AES analysis through the service of the inorganic analysis platform (PAI) at ECPM.

Reference: J. Goetz, A. Nonat, A. Diallo, M. Sy, I. Sera, A. Lecointre, C. Lefevre, C. F. Chan, K.-L. Wong, L. J. Charbonnière, *ChemPlusChem* **2016**, *81*, 526.